\documentclass[conference]{IEEEtran}
\IEEEoverridecommandlockouts
\usepackage{cite}
\usepackage{amsmath,amssymb,amsfonts}
\usepackage{algorithmic}
\usepackage{graphicx}
\usepackage{textcomp}
\usepackage{algorithm}
\usepackage{todonotes}
\usepackage{xcolor}
\def\BibTeX{{\rm B\kern-.05em{\sc i\kern-.025em b}\kern-.08em
    T\kern-.1667em\lower.7ex\hbox{E}\kern-.125emX}}
\begin{document}




\title{Dictionary Learning for Clustering on Hyperspectral Images}

\author{\IEEEauthorblockN{Joshua Bruton}
\IEEEauthorblockA{School of Computer Science and\\Applied Mathematics\\
University of the Witwatersrand\\
Johannesburg, Gauteng}
\and
\IEEEauthorblockN{Hairong Wang}
\IEEEauthorblockA{School of Computer Science and\\Applied Mathematics\\
University of the Witwatersrand\\
Johannesburg, Gauteng}}

\maketitle

\begin{abstract}
Dictionary learning and sparse coding have been widely studied as mechanisms for unsupervised feature learning. Unsupervised learning could bring enormous benefit to the processing of hyperspectral images and to other remote sensing data analysis because labelled data is often scarce in this field. We propose a method for clustering the pixels of hyperspectral images using sparse coefficients computed from a representative dictionary as features. We show empirically that the proposed method works more effectively than clustering on the original pixels. We also demonstrate that our approach, in certain circumstances, outperforms the clustering results of features extracted using principal component analysis and non-negative matrix factorisation. Furthermore, our method is suitable for applications in repetitively clustering an ever-growing amount of high-dimensional data, which is the case when working with hyperspectral satellite imagery.
\end{abstract}


\section{Introduction}

When clustering high-dimensional data we often seek an effective representation in a latent space in the pursuit of more discriminative features. Dictionary learning provides a way to learn this representation of the latent structure of the data. After a dictionary has been acquired, sparse coding allows one to reconstruct the original data using a linear combination of the dictionary atoms with a set of weight vectors called sparse coefficients. A sparse coefficient vector is higher-dimensional than the original samples in the data however they are sparse, i.e. most of the components have zero values. To accurately reconstruct an original datum, a sparse coefficient must prioritise specific atoms (columns) in the dictionary in the correct proportion. Given that the dictionary provides a general view of the data in a lower-dimensional space, the atoms that sparse coding selects to represent a particular input must provide for any variance within the mean structure of the data. Thus, as has been shown in recent literature, a sparse coefficient matrix computed using a representative dictionary learned from a specific data set will provide useful discriminative information \cite{b12, b11, b9, b6}.

In this paper, we propose a dictionary-based clustering method that takes advantage of that discriminative information. There exists a trend across many fields towards favouring unsupervised and semi-supervised techniques. The amount of data being generated in the modern world cannot be managed solely through the paradigm of supervised learning, in which a label has to be assigned to every sample. Other techniques which require less a-priori knowledge of the data are needed. Clustering, or automatically assigning labels to samples in a data set based on some similarity criteria, is one such technique. The fields of signal processing and remote sensing are no exception in this trend towards less supervised approaches. We chose to apply our method to hyperspectral images (HSIs), in particular three widely used hyperspectral data sets. Modern hyperspectral sensors have high spectral resolution covering up to several hundred bands. Hyperspectral images acquired using such sensors are high dimensional not only in the spatial domain, but also in the spectral domain. This justifies the use of feature extraction techniques that can represent this high-dimensional information in a lower-dimensional space.

Before outlining the method in detail, we summarise some of the related work in the field and explain some important background information in section II. We provide insight into parameter-tuning and implementation details in section III. In section IV, we demonstrate the efficacy of the method empirically and show that the method outperforms two simple approaches to feature extraction, principal component analysis (PCA) and non-negative matrix factorisation (NMF), in certain circumstances. We end the paper in section V with concluding remarks and describe some of the future work that could add to the results shown here.

\section{Background and Related Work} \label{back-section}

\subsection{Online Dictionary Learning and Sparse Coding} \label{dict-subsect}

In this subsection we aim to provide a working understanding of dictionary learning and of the algorithm we use to perform it. In the context of this paper, dictionaries are dense matrices of base elements that best describe the latent structure of the data from which they are learned. Dictionary learning is a ubiquitous technique with which to learn these data representations. More formally, when creating a dictionary, we aim to represent the input data in a matrix \(\textbf{X}\) with two matrices: a dictionary, \(\textbf{D}\), and a matrix of sparse coefficients, \(\textbf{A}\). Here, \(\mathbf{D}\textbf{A}\) is called the sparse approximation of \(\textbf{X}\). Basis pursuit is not a working solution, as we also require the dictionary to be overcomplete so as not to be limited by orthogonality. By ``overcomplete" we mean that if \(\mathbf{X}\in\mathbb{R}^{m\times n}\) and \(\mathbf{D}\in \mathbb{R}^{m\times k}\) then \(k>m\). The dictionary being overcomplete guarantees that, for \(\textbf{x}\in \textbf{X}\), \(\textbf{x}=\mathbf{D}\boldsymbol{\alpha}\) will have infinitely many solutions for \(\boldsymbol{\alpha}\), we evaluate these solutions based on their sparsity, this is called ``sparse coding" or ``atom decomposition"\cite{b2}. Methods for sparse coding are usually called ``pursuit algorithms" and they aim to find approximate solutions to the problem because exact computation is intractable. We perform sparse coding with a greedy pursuit algorithm called orthogonal matching pursuit (OMP) \cite{b4}.

In the problem of dictionary learning we are given \(\textbf{X}\) and aim to find \(\textbf{D}\) and \(\textbf{A}\). If we frame the problem around the solution for \(\textbf{A}\), we call this process ``simultaneous sparse approximation". We cannot approach this problem analytically, and so we turn to optimisation techniques. Optimisation techniques effectively split simultaneous sparse approximation into two manageable steps. In the first step, we compute the sparse coefficient matrix \(\textbf{A}\) given the dictionary \(\textbf{D}\); we can do this with any pursuit algorithm. In the second step we are given the sparse coefficient matrix \(\textbf{A}\) and must update \(\textbf{D}\) sensibly. An intuitive equation for performing the dictionary update is provided by \cite{b1}: 
\begin{equation}
\textbf{D}'=\underset{\textbf{D}}{\mathrm{argmin}}{1\over t}\sum_{i=1}^{t}{{1\over 2}||\textbf{x}_{i}-\textbf{D}\boldsymbol{\alpha}_{i}||_{2}^2+\lambda||\boldsymbol{\alpha}_{i}||_{1}}
\label{odl-update}
\end{equation}
where \(\textbf{D}'\) is the new dictionary, \(\textbf{x}_i\in \textbf{X}\), \(\boldsymbol{\alpha}_i\in \textbf{A}\) and \(\lambda\) is a regularisation parameter. This equation endeavours to find the dictionary \(\textbf{D}\) that minimises the reconstruction error under some constraints. The first term, minimised in isolation from the second, would find the closest approximation of \(\textbf{x}_i\) without considering the sparsity of \(\boldsymbol{\alpha}_i\). We want \(\boldsymbol{\alpha}_i\) to be sparse, and so the second term penalises \(\boldsymbol{\alpha}_i\) in proportion to the number of nonzero elements it contains\cite{b2}. This equation can be altered algebraically to instead operate on each column of the dictionary separately, this is done in Algorithm \ref{fig:o-d-alg}. We also have to ensure that the columns of \(\textbf{D}\) are normalised at the end of each update step, in practice this is enforced by normalising the columns in \(\textbf{D}\) as they are updated. The algorithm described here is called online dictionary learning (ODL) and was originally proposed in \cite{b1}. A pseudo-code implementation of ODL can be seen in Algorithm \ref{fig:o-d-alg}.

\begin{algorithm}[b]
    \caption{Online dictionary learning algorithm \cite{b1}}
    \begin{algorithmic}[1]
    \REQUIRE Number of iterations $T$,   dictionary $\textbf{D}_0$ \(\in \mathbb{R}^{m\times k}\), an algorithm, p(\textbf{x}), to draw random variable $\textbf{x}$ from data set $\textbf{X}$, regularisation term $\lambda$
    \STATE \textbf{BEGIN}
    \STATE Initialise $\textbf{A}_0\leftarrow0$ and $\textbf{B}_0\leftarrow0$
    \STATE \textbf{for} $t=1$ to $T$:
    \STATE \hspace{2mm} draw $\textbf{x}_t$ from data set using $p(\textbf{x})$
    \STATE \hspace{2mm} sparse coding step using OMP to compute $\boldsymbol{\alpha}$
    \STATE \hspace{2mm} $\textbf{A}_t\leftarrow\textbf{A}_{t-1}+\boldsymbol{\alpha}_t\boldsymbol{\alpha}_t^T$
    \STATE \hspace{2mm} $\textbf{B}_t\leftarrow\textbf{B}_{t-1}+\textbf{x}_t\boldsymbol{\alpha}_t^T$
    \STATE \hspace{2mm} \textbf{for} $i=1$ to $k$:
    \STATE \hspace{4mm} Update $i^{\text{th}}$ column of dictionary using \eqref{odl-update}
    \STATE \textbf{END}
    \end{algorithmic}
    \label{fig:o-d-alg}
\end{algorithm}

ODL is distinct from its predecessors, K-SVD\cite{b2} and the method of optimal directions (MOD)\cite{b3}, in many ways, but the most significant distinction is that at each iteration of ODL we process a \textit{single} datum and update the dictionary accordingly. ODL allows one to routinely update an already learned dictionary with new data, without reconsidering the data that has already been seen. This significantly reduces the computational cost of maintaining a representative dictionary for a regularly updated data set. This quality also allows ODL to scale more effectively to larger data sets because it allows it to take advantage of mini-batches.

With the techniques of dictionary learning and sparse coding we can compute the sparse coefficient matrix \(\textbf{A}\). \(\mathbf{D}\textbf{A}\) is a highly compressed version of the original data, and this approximation can provide high-fidelity reconstructions. Dictionary learning and sparse coding are often used for noise-removal, in-painting and signal compression. There has also been much research conducted into the discriminative power of the sparse coefficient matrix \(\textbf{A}\). Typically, this experimentation is conducted with supervised learning tasks in mind. As mentioned earlier, this work has a different focus.

\subsection{Spectral Clustering and KMeans}

Clustering is the task of automatically labelling samples such that samples with the same label are similar by some metric, each group of samples is called a cluster \cite{b7}. The samples defining the centre of each cluster are called centroids. The KMeans algorithm is a popular approach to clustering. Each sample in the data is assigned to the centroid it is most similar to. After all the assignments are complete, the centroids are updated to the centre (mean) of all the points in their respective clusters. This process is repeated until there is no change, or minimal change, in the position of each centroid. In KMeans, similarity is typically defined in terms of Euclidean distance. The simplicity of KMeans is appealing however the algorithm makes a number of limiting assumptions about the structure of the data \cite{b2}.

KMeans defines a circle around each centroid, if a sample falls within that circle it is considered a member of the cluster that that centroid defines. This means that KMeans can only define clusters that are both convex and circular. One method for tempering this assumption is applying expectation maximisation (EM) to a Gaussian mixture model. This allows for ellipsoidal clusters as well as circular ones. This technique does not scale as efficiently to high-dimensional data and so also requires the application of a feature extraction algorithm like PCA. Another limitation of this technique is that it does not address the assumption of convexity \cite{b15}.

Spectral clustering (SC) addresses the problem of convexity by clustering with respect to a mathematical graph. To achieve this we first construct a graph from our data, the graph can be constructed using any technique but KNN is certainly the most common \cite{b13}. A graph containing \(N\) vertices, corresponding to \(N\) datum \(\{S_{i}\}_{i=1}^{N}\), can be represented as an \(N\times N\) affinity matrix. In this affinity matrix each entry \((i, j)\) represents the weight of the edge between vertex \(i\) and vertex \(j\), we refer to each of these weights as \(W_{i, j}\). From an \(N\times N\) affinity matrix we can obtain \(N\) elements \(d_{i}\) with the equation \(d_{i}=\sum_{j=1}^{N}{W_{i, j}}\) for \(i\in \{1, 2, ..., N\}\), these define the diagonal entries of the degree matrix. We then construct the \(N\times N\) Laplacian matrix, \(\textbf{L}\), using the following rule \cite{b14}: 

\begin{equation}
    \textbf{L}_{i, j} = \left\{
      \begin{array}{lr}
        d_{i} & : i=j\\
        -W_{i, j} & : \text{otherwise}
      \end{array}
    \right.
    \label{laplacian}
\end{equation}

Our goal in constructing the Laplacian matrix is to perform graph partitioning, the Laplacian matrix has useful properties to facilitate this. We first decompose the Laplacian matrix into eigenvalues and eigenvectors, for an \(N\times N\) Laplacian matrix we will obtain \(N\) eigenvalues, \(\{\lambda_{1}, \lambda_2, ..., \lambda_{N}\}\), and \(N\) corresponding eigenvectors, \(\{\textbf{x}_{1}, \textbf{x}_{2}, ..., \textbf{x}_{N}\}\). The eigenvalues and eigenvectors are sorted in ascending order of eigenvalue, so \(\lambda_{1}<\lambda_{2}<...<\lambda_{N}\), this is called the ``spectrum" of the Laplacian matrix\cite{b13}. Eigenvalues appearing earlier in the spectrum correspond to points of lower connectivity in the affinity graph. Each of those eigenvalues corresponds to an \(N\)-dimensional eigenvector, and each of the \(N\) entries in each eigenvector correspond to a datum in our original data. We cluster the data using these eigenvectors. Let \(\textbf{x}_{i, j}\) represent the \(j\)-th entry in the \(i\)-th eigenvector. We can then represent each of
our datum \(S_{i}\) as \((\textbf{x}_{1,i}, \textbf{x}_{2,i}, \textbf{x}_{3,i}, ..., \textbf{x}_{N,i})\). When deciding which of the eigenvectors to use in this representation, two factors must be considered. First, if the original affinity graph was fully-connected, then the first eigenvector (corresponding to the lowest eigenvalue) need not be considered as its components are constant. Second, the number of subsequent eigenvectors (not counting the first) should be one less than the number of desired clusters\cite{b14}. Using this final representation of the data as input into the KMeans algorithm will provide a better clustering.

\subsection{Joint Sparse Representation} \label{subsect-jsr-explain}

\begin{figure}[t]
\centerline{\includegraphics[width=4cm]{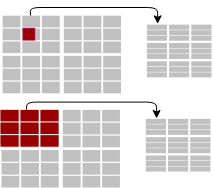}}
\caption{Standard ODL top, JSR bottom}
\label{fig:jsr-illus}
\end{figure}

As mentioned in subsection \ref{dict-subsect} and shown in Algorithm \ref{fig:o-d-alg}, ODL updates columns in the dictionary iteratively with respect to randomly selected samples. Joint sparse representation (JSR) alters this update technique to operate over a sliding window instead of a single pixel, an illustration of this can be seen in Fig. \ref{fig:jsr-illus}. This requires the dictionary to use specific columns to represent pixels that are close together in the spatial domain of the image; thus enforcing the idea of spatial locality.

Altering a standard ODL implementation to include JSR involves constructing a set of overlapping tiles, \(\{{\textbf{T}_{i}}\}_{i=1}^{N}\), each tile centred on one of the pixels in the image. Padding is not necessary, simply ignoring the borders is preferable given the size of the HSIs. A tile with dimensions \(p\) and \(q\), \(\textbf{T}_{i}\in\mathbb{R}^{p\times q}\), can be considered it's own sparse coding problem, each producing a matrix of sparse coefficients \(\textbf{Q}_{i}\in\mathbb{R}^{p\times q}\). Some pursuit algorithms have been altered to perform sparse coding on matrices, one such algorithm is simultaneous orthogonal matching pursuit (SOMP), which is outlined in \cite{b17}. The dictionary update step is also quite similar to \eqref{odl-update}, with \(\textbf{T}_{i}\) replacing \(\textbf{x}_{i}\) and \(\textbf{Q}_{i}\) replacing \(\boldsymbol{\alpha}_{i}\), and can be seen in \cite{b18}.

\subsection{Related Work}

Much research has been conducted into clustering the pixels of HSIs, although typically that research is focused on advancing clustering algorithms themselves, as opposed to using feature extraction techniques to improve existing approaches. This can be seen in \cite{b7} and \cite{b8}. The former presents an algorithm known as k Stochastic Expectation Maximisation, improving on traditional EM by selecting the number of clusters, k, using a measure of entropy instead of specifying k as a hyperparameter. The latter work has two significant contributions, DLSS (spectral-spatial diffusion learning) and active learning. DLSS is a form of clustering relying on an advanced notion of distance called diffusion distance. Active learning is essentially semi-supervised learning, in which points that are difficult for DLSS to identify can be specified before clustering begins, which results in improved accuracy. Much of \cite{b8} is devoted to expounding on these two concepts and they are out of the scope of this work.

Noting the above, some works standout as similar to ours, although none of them deal with unsupervised learning in HSIs. Reference \cite{b12} deals specifically with the use of sparse representation as a preprocessing step to broader computer vision problems, like image classification and facial recognition. Furthermore, a noteworthy conclusion drawn in \cite{b12} states that sparsity could provide a powerful prior for inference with high dimensional visual data that have intricate low-dimensional structure. This statement is exactly in line with the initial hypothesis that the work in this paper is based on. Reference \cite{b11} is also related to this research area, although that work differs from ours in some key areas. Reference \cite{b11} shows the benefits of constructing affinity matrices for SC from sparse coefficients. However, \cite{b11} does not consider learning an overcomplete dictionary as a separate problem prior to this step. One of the key findings of this work deals with the dictionary learning process specifically, and how adjustments in the hyperparameters of ODL affect clustering accuracy. This extension allows for the use of dictionary learning techniques that could further improve the construction of the affinity matrices in SC, e.g. JSR. The approach to simultaneous sparse approximation used in \cite{b11} is a technique called non-negative matrix factorisation (NMF). NMF is a powerful dimensionality reduction technique similar to dictionary learning and has been extended into a parallel dictionary learning algorithm by \cite{b16}. We will compare our method to NMF in section \ref{results-section} below.

Reference \cite{b9} also more explicitly considers dictionary learning. However, in their clustering framework, a different dictionary is trained for every cluster in the data. The similarity of a given sample to each cluster is defined by the quality of the reconstruction each of the dictionaries provides. They also define a novel way to measure the quality of the reconstruction, based on the more traditional reconstruction error defined in \eqref{odl-update}. The paper also includes the use of SC, specifically for the initialisation of the dictionaries defining each cluster. Unlike [9], we learn a unified dictionary for the entire data set as opposed to learning a different dictionary for each cluster in the data. 

\section{Proposed Method}

\subsection{Overview of the Proposed Method} \label{subsect-prop-meth}

Following the detailed background provided in section \ref{back-section}, our proposed method can be summarised in the following three steps:
\begin{enumerate}
    \item Dictionary learning: in this step, given a hyperspectral image, \(\mathbf{X}\in\mathbb{R}^{m\times n}\), we aim to learn a representative dictionary \(\mathbf{D}\in \mathbb{R}^{m\times k}\). This can be achieved using any dictionary learning technique but we proceed by using ODL, shown in Algorithm \ref{fig:o-d-alg}.
    \item Sparse coding: once the dictionary \(\mathbf{D}\in\mathbb{R}^{m\times k}\) has been found, we can proceed to find the sparse coefficient set \(\mathbf{A}\in\mathbb{R}^{k\times n}\) such that \(\mathbf{X}\approx \mathbf{D}\mathbf{A}\); which means for all \(\textbf{x}\in \mathbf{X}\) we have a sparse coefficient \(\boldsymbol{\alpha} \in \mathbf{A}\) such that \(\textbf{x}\approx \mathbf{D}\boldsymbol{\alpha}\).
    \item Clustering: We now apply a clustering technique to the sparse coefficients \(\textbf{A}\in\mathbb{R}^{k\times n}\) found in step 2. We have had the most success using SC, we construct the affinity matrix using KNN and then cluster the resulting eigenvectors using KMeans. We also present some results obtained by applying KMeans directly to \(\textbf{A}\).
\end{enumerate}

The method can be evaluated by comparing the resultant clustering after step 3 to a basic clustering method, in which the hyperspectral image clustering is carried out in the original pixel domain instead of on the sparse coefficient vectors. Comparing cluster accuracy is a non-trivial task, given that the numerical value representing each label is assigned arbitrarily, a simple element-wise comparison of a clustering with the ground truth is meaningless. Although visually assessing the accuracy of a particular clustering is possible and oftentimes simple, having a more consistent measure is desirable. Two popular metrics for this task are AMI (Adjusted Mutual Information score) and ARI (Adjusted Rand Index). According to \cite{b5}, when comparing clusters on data with unbalanced or small clusters, one should prioritise the AMI. Given that HSIs typically have these qualities, we have chosen to use AMI to evaluate the method. 

AMI will return \(1.0\) if the clustering is perfect, the average return for entirely random clusterings is \(0\), this means AMI can also be negative. The procedure for computing the AMI requires both the ground-truth and the clustering being evaluated, so labelled data is required. AMI is an extension of an earlier score for measuring the accuracy of clusterings, called mutual information (MI) score. MI attempts to quantify the similarity between two partitions, it does this probabilistically. We are given two equally-sized partitions with the same number of clusters in each, we denote the number of clusters as k and the total number of datum as \(n\). Then, let \(\{G_{i}\}_{i=1}^{k}\) and \(\{L_{i}\}_{i=1}^{k}\) denote the k clusters in each of our partitions; in our case, the ground-truth and predicted labels, respectively. We can now find the probability that a point drawn at random from each partition is associated with a particular cluster, \(i\), using \(\textbf{P}(i)={|G_{i}|\over{n}}\) and \(\textbf{P}'(i)={|L_{i}|\over{n}}\). We can also compute the probability that a random point is in cluster \(i\) in the ground truth and cluster \(j\) in the prediction by evaluating \(\textbf{P}(i,j)={{|G_{i}\cap L_{j}|}\over{n}}\). With these expressions, we can define the mutual information of partitions G and L as:

\begin{equation}
    MI(G, L)=\sum_{i=1}^{k}{\sum_{j=1}^{k}{\textbf{P}(i,j)\log{{\textbf{P}(i,j)\over{\textbf{P}(i)}\textbf{P}'(j)}}}}
    \label{mutual-info}
\end{equation}

Given this formulation of MI we also know that the entropy of partitions \(G\) and \(L\) are given by \(E(G)=-\sum_{i=1}^{k}{\textbf{P}(i)\log{\textbf{P}(i)}}\) and \(E(L)=-\sum_{i=1}^{k}{\textbf{P}'(i)\log{\textbf{P}'(i)}}\), respectively. As per \cite{b5}, extending \eqref{mutual-info} and incorporating entropy, the AMI of partitions G and L is defined as follows: 

\begin{equation}
    AMI(G, L)={ {MI(G, L)-\mathbb{E}[MI(G, L)]}\over{\max{[E(G), E(L)]}-\mathbb{E}[MI(G, L)]} }
    \label{adjusted-mutual-info}
\end{equation}

The most important distinction between MI and AMI is that AMI incorporates the idea of chance. The score that a clustering receives from AMI will be reduced in proportion to how easily its MI score could be produced by a random clustering, this is encapsulated in \(\mathbb{E}[MI(G, L)]\), the expected MI. AMI is similar to the popular Cohen's kappa coefficient (widely used in supervised learning), both endeavour to incorporate the percentage agreement and the possibility of that percentage agreement occurring by chance.

\subsection{Hyperparameters}

The proposed method has four hyperparameters that require tuning. In this subsection we will discuss the affects of these hyperparameters. In the first step we perform ODL, any pursuit algorithm can be used; we elected to use OMP so we could select exactly how many non-zero elements to extract in the sparse coding step. When performing dictionary learning you must specify the number of atoms in the dictionary, the regularisation parameter, \(\lambda\), and the number of iterations, \(T\). 

The number of atoms in the dictionary must be greater than the dimensionality of the input signals, we found a reasonable heuristic for the number of atoms is double the dimensionality of the signals. \(\lambda\) specifies how heavily to prioritise the sparsity of the sparse coefficients, higher \(\lambda\) values correspond to greater sparsity. In general, we require the sparse coefficient vectors to be as sparse as possible while still providing for accurate data reconstruction. Finally, \(T\) specifies the number of iterations over which to learn the dictionary; \(T\) is particularly data specific but ODL guarantees convergence on a specific reconstruction error past-which further training is unnecessary.

\section{Experiments and Results} \label{results-section}

\subsection{Domains}

\begin{figure}[t]    
\centerline{\includegraphics[width=8cm]{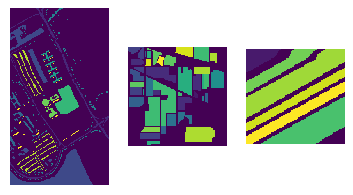}}
\caption{Left to right: ground-truth of Pavia University, Indian Pines and Salinas}
\label{fig:data-display}
\end{figure}

In this section we present our experimental results. We ran experiments on three widely used HSIs, namely: Salinas, Indian Pines and Pavia University\cite{b10}, the ground truth labels of each of these images can be seen in Fig. \ref{fig:data-display}. HSIs are images with a large number of spectral channels, typically over one-hundred. An HSI{'}s information can be seen as operating in a 3-dimensional cube, where along the two spatial dimensions it is simply a series of 2-dimensional images, and along the third (or spectral) dimension, it consists of images taken at various wavelengths. In this structure, each pixel in the hyperspectral image can be represented as a continuous signal; this makes HSIs ideal candidates for signal processing techniques. In this work, we do not take advantage of the spatial information within the HSIs, so the method can be extended to other domains within signal processing.

Although these three data sets are all HSIs, they each present their own challenges. Salinas, captured by the AVIRIS spectrometer, is \(83\times 86\) and contains \(204\) spectral bands. It contains six classes and one miscellaneous class, given its simplicity, high-accuracy predictions are expected. Indian Pines, also captured by AVIRIS, contains sixteen different classes and its dimensions are \(145\times 145\times 200\), it is considered far more challenging than Salinas\cite{b10}. Since both Salinas and Indian Pines were captured by AVIRIS, they have similar spectral resolutions. Pavia University, captured by ROSIS, has only \(103\) spectral bands, however, at \(610\times 340\), it is much larger than the other two images. Pavia University has nine classes and so is between Salinas and Indian Pines in terms of label-complexity. These three data sets are used in the field for a wide-range of experiments, as can be seen in \cite{b6, b7} and \cite{b8}.

\subsection{Methodology}

Our Python implementation of the method described in subsection \ref{subsect-prop-meth} will be made available as of \(18\) October 2019\footnote{https://github.com/JoshuaDBruton/SparseCoefficientClustering}. The data sets used in our experiments are available as .mat files on the Grupo de Inteligencia Computacional website\footnote{http://tiny.cc/f6gnez}. As is common when working with these HSIs, we ignored the miscellaneous classes when clustering. All of the experiments described in this paper were validated across all three images.

We began our experiments by training dictionaries, these dictionaries were saved as .npy files and used for clustering at a later stage. This means that the same dictionaries could be used when comparing different clustering techniques. When clustering, the number of clusters was set to the number of classes present in the ground-truth. For the experiments shown here, we used OMP as the pursuit algorithm. OMP's stopping criteria was always the number of non-zero elements in the coefficients, this allows for fairer comparison to other feature extraction techniques, like PCA, in which the number of components is specified.

When describing our experiments we require some terms that are not available succinctly in the field, and so we will elaborate on their meaning here. By ``discriminative dictionary" we mean a dictionary whose sparse coefficients provide features that result in adequate clustering accuracy. By ``size" of the dictionary we mean the number of atoms specified for the dictionary. We term the number of non-zero elements targeted in the sparse coding step of ODL the ``sparsity of the dictionary".

\subsection{Sparsity and Number of Components} \label{subsect-find-dict}

\begin{figure}[t]    
\centerline{\includegraphics[width=8cm]{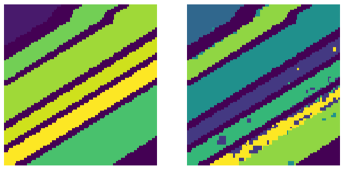}}
\caption{Ground-truth left, prediction of KMeans on original pixels right, colour of classes is arbitrary, note significant error in bottom-right class}
\label{fig:KM-baseline}
\end{figure}

We first undertook the task of establishing the correct size and sparsity to target when creating a discriminative dictionary. We provide a baseline clustering for Salinas in Fig. \ref{fig:KM-baseline}, this baseline was created by applying KMeans to the orignal pixels of Salinas. Since the intention of this experiment was to compare the discriminative power of the dictionaries, our only concerns were the relative performance of the final sparse coefficient predictions and how they compared to this baseline. This also justifies our use of KMeans instead of a more complex clustering technique.

We proceeded with a grid-search over some possible parametrisations of ODL. We trained 16 different dictionaries for each HSI, each with various sizes and sparsities. These dictionaries and some visualisations of the training process are available on Comet.ml\footnote{https://www.comet.ml/joshuabruton/honours-project/view/}. We then performed sparse coding using these dictionaries, and used the computed sparse coefficients as inputs to KMeans. We displayed the resultant predictions in a grid to see if any obvious patterns emerged.

\begin{figure}[t]    
\centerline{\includegraphics[width=6cm]{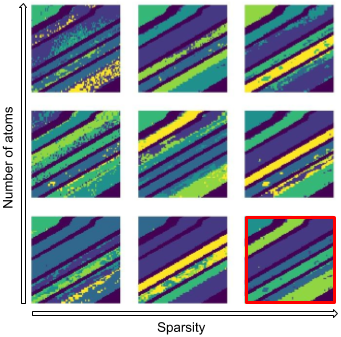}}
\caption{Grid-search over ODL hyperparameters, increasing sparsity on x-axis, increasing size on y-axis, the best result is highlighted in the bottom-right}
\label{fig:gridsearch}
\end{figure}

A \(3\times 3\) sample of the grid created for Salinas can be seen in Fig. \ref{fig:gridsearch}. Nine dictionaries are compared in the grid, all were trained for \(5000\) iterations. The number of atoms in each dictionary varies over \((220, 306, 408)\) and the sparsity varies over \((10, 5, 2)\). Every combination of those values is shown in the grid, with the smallest and sparsest dictionary in the bottom-right and the largest and densest dictionary in the top-left.

As can be seen in the grid, the sparsest and smallest dictionary, targeting \(2\) non-zero elements in sparse coefficients with size \(220\), provided the best prediction. This prediction is similar to the baseline prediction provided in Fig. \ref{fig:KM-baseline} and many of the errors are common to both predictions. This result was not anticipated, the denser and larger dictionaries have substantially lower reconstruction errors. We initially expected lower reconstruction error to correlate with predictions more similar to the baseline. In hindsight, this result confirms the idea that the sparse coefficients intend to represent the variance in the signals while the dictionary represents their generic structure. It also confirms that sparse coding prioritises the extraction of more discriminative features. Making the sparse coefficients too large, or too dense, effectively allows OMP to represent any noise within the signals.

\subsection{Joint Sparse Representation}

\begin{figure}[b]    
\centerline{\includegraphics[width=8cm]{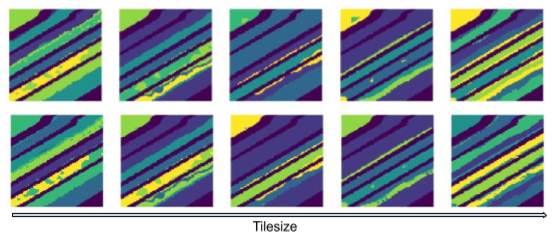}}
\caption{Mean of tile top, middle pixel of tile bottom}
\label{fig:jsr-comparison}
\end{figure}

Before proceeding with experiments relating to SC, we tested the performance of the most trivial implementation of JSR, as outlined in \ref{subsect-jsr-explain}, with KMeans. We trained five dictionaries using JSR, all of them had size \(330\) and targeted \(2\) non-zero elements in the sparse coding step. The tile-sizes of the dictionaries varied over \((3, 5, 7, 9, 11)\). When using JSR, after the sparse coding step, each pixel is represented by a tile of sparse coefficients. We thought of two approaches for reducing this tile to one datum so it could be clustered. Either taking the mean of the entire tile, or simply extracting the centre sparse coefficient from the tile. We attempted both approaches, they have similar results and a comparison can be seen in Fig. \ref{fig:jsr-comparison}.

The predictions provided by JSR are not as accurate as we hypothesised. JSR's predictions seem to be less accurate than the KMeans baseline. There are more complex approaches to JSR that may have produced better results, some of which are described in \cite{b19}. However, most of these alternatives are supervised, placing them out of the scope of this work.

\subsection{Feature extraction}

We now examine the benefit of using the most discriminative dictionaries we found in experiment \ref{subsect-find-dict} to produce sparse coefficients as input to SC. The intention is that these sparse coefficients will allow for more effective construction of the affinity matrices. We compared the results of four approaches to spectral clustering; the first using the original signals, the second using principal components from PCA, the third using non-negative matrices computed from NMF, and the fourth using the sparse coefficients computed from dictionaries. For this experiment, the three smallest and sparsest dictionaries found in the previous experiment were used, one for each HSI. All three targeted a sparsity of \(5\) in the sparse coding step that precedes training. The dictionaries trained on Salinas and Indian Pines both had \(220\) atoms and the dictionary trained on Pavia had \(120\) atoms. The simplicity of ODL's hyperparameter selection is an important characteristic of the method. We used PCA to extract \(50\) principal components and clustered on those components. We provide results for NMF with 8 components, a higher number tended to provide grainy predictions. It is worth noting that both NMF and PCA use larger representations than the one used for the sparse coefficients, PCA significantly so.

\begin{figure}[b]    
\centerline{\includegraphics[width=9cm]{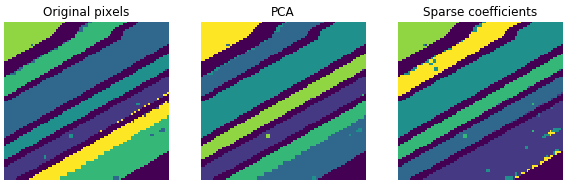}}
\caption{Left to right: SC on original pixels, SC on 50 principal components, SC with 5 non-zero elements}
\label{fig:final-comparison}
\end{figure}

The results of this comparison are shown in Table \ref{table:pcaodl}. SC generally performs poorly on the original pixels, this is because constructing the affinity matrix from high-dimensional data is ineffective. PCA and sparse coding effectively compensate for this by creating less complex but still meaningful representations. The effect of sparse coding is less noticeable on Pavia University, we conclude that this is because it has the fewest spectral channels. Notably, spectral clustering on the sparse coefficients works most effectively on the two images with the largest number of channels. Dictionary learning and sparse coding are designed specifically for application to high-dimensional signals, this explains the under-performance of PCA and NMF in these cases. Relative to PCA and the sparse coefficients, NMF performs poorly on the two images with the highest dimensional pixels but performs best on Pavia University. The improvement our method provides is especially noticeable on Salinas, with a \(0.0399\) increase in the AMI. SC is well-suited to Salinas, however, only the sparse coefficients provide features powerful enough to take advantage of this compatibility. A visual comparison of the top three Salinas predictions is shown in Fig. \ref{fig:final-comparison}.

\begin{table}[t]
\caption{Comparison of different features in spectral clustering}
\begin{center}
\begin{tabular}{l c c c c}
\hline
\multicolumn{5}{c}{\textbf{Adjusted mutual information score$^{\mathrm{a}}$}} \\
\cline{1-5} 
\textbf{HSI} & \textbf{\textit{Original pixels}}& \textbf{\textit{PCA}} & \textbf{NMF} & \textbf{\textit{Sparse coefficients}} \\
\hline
Salinas & 0.8981 & 0.9088 & 0.8731 & \textbf{0.9487} \\
\hline
Indian Pines & 0.6310 & 0.6489 & 0.626 & \textbf{0.6509} \\
\hline
Pavia U & 0.8056 & 0.83 & \textbf{0.849} & 0.8065  \\
\hline
\multicolumn{5}{l}{$^{\mathrm{a}}$An AMI of 1.0 signifies a perfect clustering, higher is better}
\end{tabular}
\label{table:pcaodl}
\end{center}
\end{table}

\subsection{Discussion}

Our experiments here show that it is possible to represent HSIs in the significantly compressed form provided by ODL while still maintaining enough information to cluster as or more effectively than in the original hyperspectral pixel domain. Clustering the sparse coefficients is also more computationally efficient as far fewer comparisons are needed. This computational efficiency will be cancelled out if clustering is not performed frequently, but could be beneficial when clustering similar data repetitively. The benefit of using ODL in this scenario is that the same dictionary can be continually updated as new data is added, without reconsidering what data was previously seen. As has been shown by \cite{b2}, dictionary learning does allow for generalisability. This means it would also be possible to generalise a dictionary trained on a similar data set to other datum, meaning that a new dictionary does not have to be found for every new clustering problem. We also intend on further investigating joint sparse representation. By taking advantage of the available spatial information, joint sparse representation should greatly improve spectral clustering accuracy in HSIs.

\section{Conclusion}

In this paper, we proposed a method for clustering hyperspectral images via dictionary learning and sparse coding. In particular, we investigated a number of different clustering approaches for hyperspectral images. This includes clustering in the original pixel domain, clustering in the sparse coefficient domain, spectral clustering in the sparse coefficient domain, spectral clustering using non-negative matrix factorisation, and clustering in the PCA coefficient domain. Our experimental results show that using sparse coefficients led to competitive or superior performance when compared to the other methods considered in this paper.


\begin{thebibliography}{00}
\bibitem{b1} J. Mairal, F. Bach, J. Ponce, and G. Sapiro, ``Online dictionary learning for sparse coding,'' $26^{th}$ International Conference on Machine Learning, Montreal, pp. 689--696, 2009.
\bibitem{b2} M. Aharon, M. Elad, and A. Bruckstein, ``{K-SVD}: An algorithm for designing overcomplete dictionaries for sparse representation," IEEE Transactions on Signal Processing, vol. 54, pp. 4311--4322, November 2006.
\bibitem{b3} K. Engan, S. O. Aase, and J. H. Hus{\o}y, ``Multi-frame compression: theory and design," EURASIP Signal Process, vol. 80, pp. 2121--2140, 2000.
\bibitem{b4} S. Mallat and Z. Zhang, ``Matching pursuits with time-frequency dictionaries," IEEE Trans. Signal Processing, vol. 41, pp. 3397--3415, 1993.
\bibitem{b5} S. Romano, N. Vinh, J. Bailey, and K. Verspoor, ``Adjusting for chance clustering comparison measures," Journal of Machine Learning Research, vol. 17, pp. 1--32, 2016.
\bibitem{b6} H. Wang and T. Celik, ``Sparse representation-based hyperspectral image classification," Signal, Image and Video Processing, vol. 12, pp. 1009--1017, July 2018.
\bibitem{b7} C. Cariou and K. Chehdi, ``Unsupervised nearest neighbors clustering with application to hyperspectral images," IEEE Journal of Selected Topics in Signal Processing, vol. 9, no. 6, pp. 1105--1116, September 2015.
\bibitem{b8} J. Murphy and M. Maggioni, ``Unsupervised clustering with active learning of hyperspectral images with nonlinear diffusion," IEEE Transactions on Geoscience and Remote Sensing, vol. 57, no. 3, pp. 1829--1845, 2018.
\bibitem{b9} P. Sprechmann and G. Sapiro, ``Dictionary learning and sparse coding for unsupervised clustering," IEEE international conference on acoustics, speech and signal processing, pp. 2042--2045, 2010.
\bibitem{b10} R. Green, M. Eastwood, C. Sarture, T. Chrien, M. Aronsson, B. Chippendale, J. Faust, B. Pavri, C. Chovit, and M. Solis, et al., ``Imaging spectroscopy and the airborne visible/infrared imaging spectrometer (AVIRIS)," Remote sensing of environment, vol. 65, no. 3, pp. 227--248, 1998.
\bibitem{b11} X. Feng, ``Robust spectral clustering via sparse representation," Recent Applications in Data Clustering, p. 155, August 2018.
\bibitem{b12} J. Wright, Y. Ma, J. Mairal, G. Sapiro, T. Huang, and S. Yan, ``Sparse representation for computer vision and pattern recognition," Proceedings of the IEEE, vol. 98, no. 6, pp. 1031--1044, 2010.
\bibitem{b13} Y. Weiss, ``Segmentation using eigenvectors: a unifying view," Proceedings of the seventh IEEE international conference on computer vision, vol. 2, pp. 975--982, 1999.
\bibitem{b14} A. Ng, M. Jordan, and Y. Weiss, ``On spectral clustering: analysis and an algorithm," Advances in neural information processing systems, pp. 849--856, 2002.
\bibitem{b15} C. Bishop, ``Pattern recognition and machine learning," Springer, 2006.
\bibitem{b16} Z. Tang, S. Ding, Z. Li, and L. Jiang, “Dictionary learning based on nonnegative matrix factorization using parallel coordinate descent,” Abstract and Applied Analysis, vol. 2013, Article ID 259863, 2013.
\bibitem{b17} J. Tropp, A. Gilbert, and M. Strauss, ``Algorithms for simultaneous sparse approximation. Part I: greedy pursuit," Signal Processing,vol. 86, no. 3, pp. 572--588, Elsevier, 2006.
\bibitem{b18} J. Mairal, F. Bach, J. Ponce, and G. Sapiro, ``Online learning for matrix factorization and sparse coding," inria-00408716v1, 2009.
\bibitem{b19} Z. Wang, J. Liu, and J. Hue, ``Joint sparse model-based discriminative K-SVD for hyperspectral image classification," Signal Processing, vol. 133, pp. 144--155, Elsevier, 2017.
\end{thebibliography}
\end{document}